# Giant photo-effect in proton transport through graphene membranes

Marcelo Lozada-Hidalgo[1], Sheng Zhang[1], Sheng Hu[2], Vasyl G. Kravets[1], Francisco J. Rodriguez[1], Alexey Berdyugin[1], Alexander Grigorenko[1], Andre K. Geim[1]

*[1]School of Physics & Astronomy, University of Manchester*
*Manchester M13 9PL, UK*
*[2]National Graphene Institute, University of Manchester*
*Manchester M13 9PL, UK*

**Graphene has recently been shown to be permeable to thermal protons[1], the nuclei of hydrogen atoms, which sparked interest in its use as a proton-conducting membrane in relevant technologies[1-4]. However, the influence of light on proton permeation remains unknown. Here we report that proton transport through Pt-nanoparticle-decorated graphene can be enhanced strongly by illuminating it with visible light. Using electrical measurements and mass spectrometry, we find a photoresponsivity of ~$10^4$ A $W^{-1}$, which translates into a gain of ~$10^4$ protons per photon with response times in the microsecond range. These characteristics are competitive with those of state-of-the-art photodetectors that are based on electron transport using silicon and novel two-dimensional materials[5-7]. The photo-proton effect can be important for graphene's envisaged use in fuel cells and hydrogen isotope separation. Our observations can also be of interest for other applications such as light-induced water splitting, photocatalysis and novel photodetectors.**

Recent experiments have established that graphene monolayers are surprisingly transparent to thermal protons, even in the absence of lattice defects[1,2]. The proton transport through graphene was found to be thermally activated[1] with a relatively low energy barrier of about 0.8 eV. Further measurements involving hydrogen's isotope deuterium have shown that this barrier is in fact 0.2 eV higher than the measured activation energy because the initial state of incoming protons is lifted by zero-point oscillations at oxygen bonds within the proton-conducting media used in the experiments[2]. The resulting value of ~1.0 eV for the graphene barrier is somewhat lower (by at least 30%) than the values obtained theoretically for ideal graphene[1,8-10], which triggered a debate about the exact microscopic mechanism behind the proton permeation[8-12]. For example, it was recently suggested that graphene's hydrogenation could be an additional ingredient involved in the process[11]. Independently of fundamentals of the involved mechanisms, the high proton conductivity of graphene membranes combined with their impermeability to other atoms and molecules entices their use for various applications including fuel cell technologies and hydrogen isotope separation[1-4]. For example, it was argued that mass-produced membranes based on chemical-vapor-deposited graphene can dramatically increase efficiency and decrease costs of heavy water production[1,2,4].

In this Letter, we describe an unexpected enhancement of proton transport through catalytically-activated[1] graphene under low-intensity illumination. The devices used in this work were made from monocrystalline graphene obtained by mechanical exfoliation. The graphene crystals were suspended over microfabricated holes (~10 $\mu$m diameter) etched in silicon-nitride films, following the recipe reported previously[1,2]. The resulting free standing membranes were decorated on one side with Pt nanoparticles deposited using electron-beam evaporation (Fig. 1a). On the opposite side of the free standing membrane, a proton-conducting polymer (Nafion[13]) was drop cast and then contacted with a proton-injecting electrode[1]. In this setup, if a negative bias is applied to graphene, protons are injected into the Nafion film and then pass through the graphene membrane, evolving into $H_2$ on the side decorated with Pt nanoparticles[1,2]. Graphene – a mixed electron-proton conductor – acts here as both proton conducting membrane and cathode. Further details of device fabrication and electrical measurements are provided in Supplementary Information (see Supplementary Fig. 1)

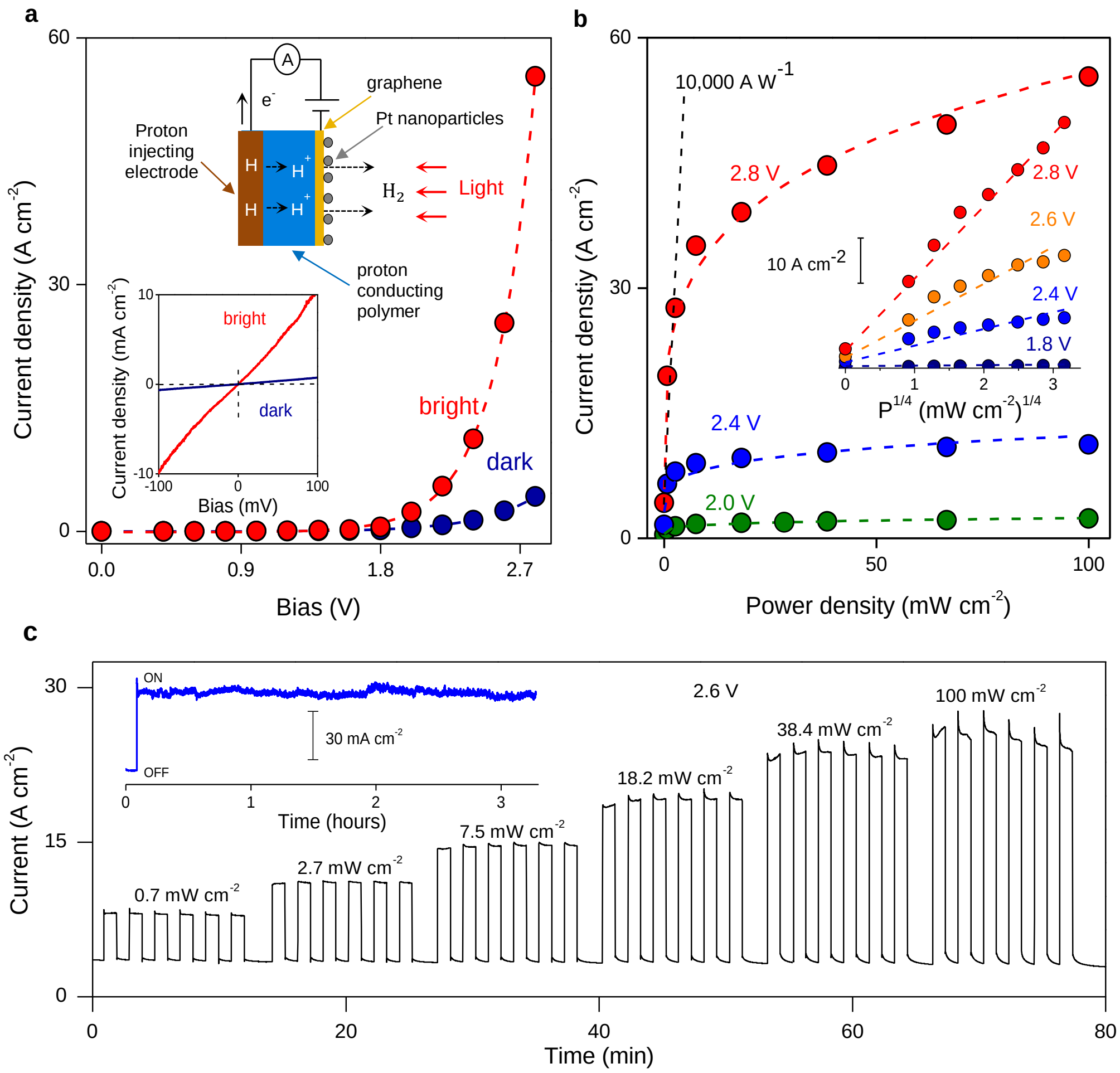

**Figure 1| Influence of illumination on proton transport through graphene activated with Pt nanoparticles. a**, Current-voltage characteristics for one of our devices under dark and bright conditions. Dashed curves, Guides to the eye. Top inset, Schematic of our measurement setup. Bottom inset, Photo-proton effect at low biases. **b**, Proton current *I* as a function of illumination power *P* for different biases. Dashed curves, Guides to the eye. The straight black line indicates responsivity at low illumination powers. Inset: Photo-proton effect can be described by the dependence $I \propto P^{1/4}$. **c**, Changes in *I* induced by one minute long illumination at a bias of 2.6 V; six measurements for each power density. The inset shows that the bright current remained stable under continuous illumination; voltage bias 0.4 V.

Figure 1a shows typical current-voltage (*I-V*) characteristics of our devices measured in the dark and under simulated solar illumination of 100 mW $cm^{-2}$ (using light source Oriel Sol3A). We found that the *I-V* response during illumination scaled by a factor of ~10 with respect to the response in the dark. Indeed, at low biases, where the *I-V* characteristics were linear, the slope of *I* increased by a factor of ~10 with respect to the dark case (inset Fig. 1a). And the same enhancement of the current by illumination was observed in the high-*V* regime, where *I-V* characteristics are nonlinear (see Fig. 1a and Supplementary Fig. 2). Further measurements showed that the current density (*I*) displayed a saturating dependence as a function of the illumination power density (*P*). This dependence could be accurately described empirically using the relation $I \propto P^{1/4}$ (see Fig. 1b). Interestingly, at low illumination ($P < 5$ mW $cm^{-2}$), our data can also be approximated by a linear dependence on *P*, which yields a photoresponsivity of ~$10^4$ A $W^{-1}$ and, probably, even higher at lower intensities (<1 mW $cm^{-2}$) because of the nonlinear functional form *I*(*P*). For white light, this translates into a large gain of ~$10^4$ protons per incident photon.

The exceptionally high photoresponsivity of our devices calls for their further characterization. First, the bright current did not show any sign of deterioration under long continuous illumination (inset Fig. 1c). Furthermore, some devices were retested after several months and displayed the same photo-proton effect. Second, using chopped illumination (bottom inset of Fig. 2), we observed that changes in *I* were fast being limited by the temporal resolution of our experimental setup. This allowed only an upper bound estimate of the intrinsic response time as ≤ 50 μs. Furthermore, these measurements showed that devices were stable and allowed at least ~$10^6$ ON-OFF cycles. Third, using different wavelength filters, we found that the devices displayed a flat response with no spectral features within the studied range of 450 nm to 1480 nm (Supplementary Fig. 3). Fourth, we determined the devices' noise equivalent power (*NEP*) within their operational range up to 2.8 V. This figure of merit describes the minimum radiant power measurable, taking into account the finite electric noise in the dark. We found *NEP*~$10^{-14}$-$10^{-16}$ W $Hz^{-1/2}$ (Supplementary Fig. 4), which indicates high sensitivity. See "Figures of merit" in Supplementary Information.

It is instructive to compare the above characteristics due to the photo-proton effect with those of photodetectors based on electron transport. Three parameters are usually used to evaluate photodetectors: photoresponsivity, photodetectivity and response time. Let us look first at graphene and other two-dimensional crystals, materials that attract intense interest in many photo-electronic applications such as phototransistors, terahertz detectors and bolometers[5,6,14]. Without adding extra photosensitive ingredients (such as, e.g., quantum dots), graphene photodetectors typically exhibit photoresponsivities[5] ranging between ~0.1 and 100 mA $W^{-1}$. Other 2D crystals can exhibit a much stronger response[5,6,14]. A prominent example is monolayer $MoS_2$, which shows[14] a photoresponsivity of ≈880 A $W^{-1}$ and a *NEP* of ~$10^{-15}$ W $Hz^{-1/2}$, albeit with a slow rise time of ≈4 s. Another pertinent reference is commercial silicon photodiodes[7,15]. Those typically display[7,16] a photoresponsivity ~100 mA $W^{-1}$, a *NEP* of ~$10^{-12}$ W $Hz^{-1/2}$ and a response time around ~0.01-$10^3$ μs. In comparison, the photoresponsivity of our devices is at least 10 times higher than the value recently highlighted for $MoS_2$ photodetectors[14] and $10^5$ times higher than for commercial photodiodes. The response time, although currently limited by our measurement setup, is at least $10^5$ times faster than for $MoS_2$ and comparable to non-specialized silicon photodiodes. The third parameter, *NEP*, is comparable or higher to that of $MoS_2$ and commercial silicon photodetectors. This comparison suggests that the proton-based devices may be suitable for some applications even in their current non-optimized design, especially if all figures of merit are considered together.

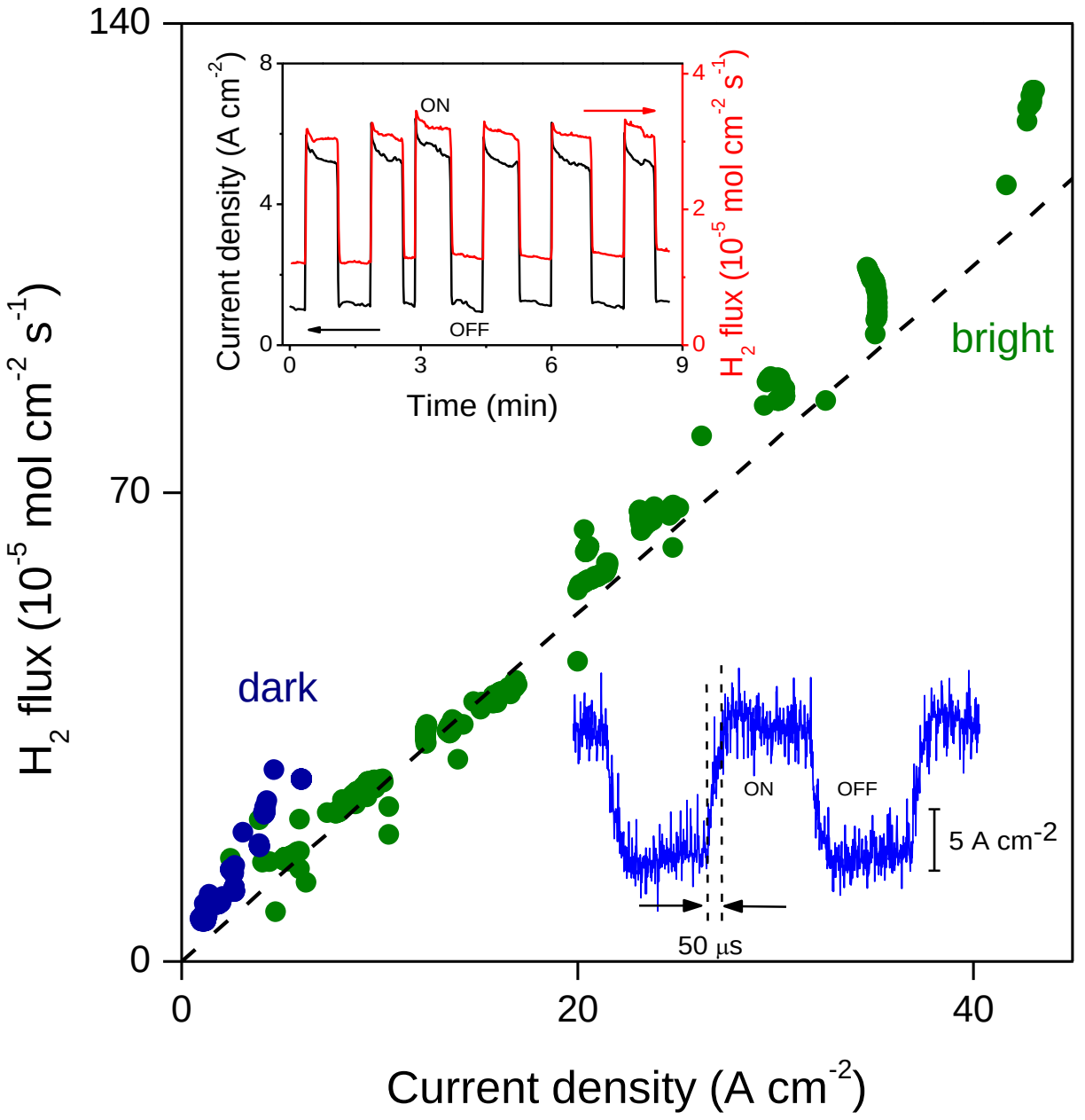

**Figure 2| Photo-proton effect observed by mass spectrometry and micro-scale time response.** Hydrogen flux and current density under dark and bright conditions, using biases in the range from 0 to 3 V. Dashed line, Faraday's law. Top inset: Example of raw data for simultaneously recorded $I$ and $\Phi$ while switching illumination on and off (black curve, Current; red, Hydrogen flux) Bias, 2.3 V. Bottom inset: Frequency response for the proton current measured using 1 kHz chopped illumination at 2.8 V.

To gain more insight into the photo-proton effect, we studied it by measuring the proton flux directly rather than using the electric current as a proxy. The discussed devices (Fig. 1a) were placed to separate two chambers, one of which contained an $H_2O/H_2$ gas mixture and the other was evacuated and connected to a mass spectrometer[1,2]. The graphene membrane served as a cathode and faced the spectrometer chamber. The proton-injecting electrode worked as an anode and faced the gas chamber. If no bias was applied or if the voltage polarity was reversed, no hydrogen flow could be detected, as expected because of graphene's impermeability to gases[1,17]. For the proper polarity, both electrical current and $H_2$ flow were detected and recorded simultaneously. The device worked effectively as an electrochemical pump demonstrating a 100% Faradaic efficiency[1,2]. The latter means that for every two electrons that flowed through the electrical circuit, an $H_2$ molecule appeared in the vacuum chamber. This charge-to-mass conservation is described by Faraday's law of electrolysis: $\Phi = I/2F$, where $\Phi$ is the hydrogen flux, $I$ is the current density, $F$ is Faraday's constant and the factor of 2 accounts for the two protons required to form a hydrogen molecule[1,2]. In the context of this report, if the devices were illuminated at a given bias, we observed a simultaneous increase in the current density and the hydrogen flux. The Faradaic efficiency under illumination remained at 100% (Fig. 2). These results corroborate those from our electrical measurements. It is instructive to note that to generate a mole of hydrogen, our electrochemical pumps require the energy input $E = IV/\Phi = 2FV$, which at typical biases translates into ~50 W h per g of hydrogen.

To narrow down the number of possible mechanisms behind the photo-proton effect we performed some additional measurements. First, temperature dependence measurements showed that the dark current displayed an Arrhenius type behavior: $I \propto \exp(-E/k_B T)$, where $E$=0.4±0.06 eV is the activation energy, $k_B$ is the Boltzmann constant and $T$ is the temperature (Supplementary Fig. 2). Within the resolution of our measurement, ±60 meV, we did not observe a significant change in $E$ under illumination. Second, the photo-proton effect was also observed for graphene membranes decorated with other catalytically-active metals[18-20] such as Pd and Ni. In both cases, we observed the same power dependence $I \propto P^{1/4}$ as for Pt but the effect was a factor of ~2 weaker for both Pd and Ni (Supplementary Fig. 2). We note that Pd, Pt and Ni are known to strongly interact with graphene[18-21] and provide n-type (electron) doping[18-20]. In contrast, if we used nanoparticles of Au, a metal that weakly interacts with graphene and p-dopes it[18,20], the proton current remained unaffected even by our strongest illumination (100 mW cm$^{-2}$). Hence, the mechanism behind the photo-proton effect must be consistent with the following observations. The effect is observed only for metals that n-dope graphene; the photo-response saturates as $I \propto P^{1/4}$; it is stable for many hours of continuous illumination; the activation energy for the process does not change noticeably under illumination, and the effect is present even in the linear regime, at very small biases. With these facts in mind, we considered a number of possible scenarios for the photo-effect. A reduction in the transport barrier due to plasmonic effects that explain many photochemical reactions can be ruled out because of the lack of response in the presence of plasmon-active Au nanoparticles[22,23]. Photo-induced electrolysis of water is also readily ruled out because the photo-proton effect appears at low $V$, well below the thermodynamic voltage for water splitting and its relative amplitude changes little with $V$. Even if the electrolysis did occur, its products cannot possibly cross through our membranes that are impermeable to hydrogen, even in its atomic form[8,9,17]. Photo-induced hydrogenation of graphene can in principle reduce the energy barrier[10,11] but in this case the illumination would also be expected to change $E$, in contrast to the experiment.

After ruling out the above and some other deemed mechanisms, we suggest the following explanation. Metal nanoparticles are known to dope graphene underneath, which leads to an in-plane electric field in the graphene areas surrounding them[18,20]. It is also known that illumination of such built-in junctions creates long-lived (>1 ps) hot electrons in graphene[24,25], and these electrons generate a photovoltage[5,26,27], similar to illumination of semiconducting p-n junctions. The photovoltage $V_{ph}$ in graphene is proportional to its electron temperature, $T_e$. We believe that this local voltage influences proton permeation in the following way. If we use Pd, Ni or Pt nanoparticles – metals that dope graphene with electrons – the in-plane electric field attracts photo-generated electrons towards the nanoparticles and holes away from them. Hence, similar to a negative voltage applied to the whole graphene membrane, this local voltage $V_{ph}$ should also funnel protons and electrons towards the nanoparticles. This results in an enhanced rate of electron-proton conversion into atomic hydrogen. The photovoltage is able to influence the latter rate because the attempt rate for proton permeation through graphene is of the order[2,28] of $10^{13}$ - $10^{14}$ s$^{-1}$; that is ~100 times faster than the lifetime of hot electrons. In contrast, if we use nanoparticles of Au (a metal that dopes graphene with holes), photo-generated electrons move in the opposite direction, and a positive $V_{ph}$ appears at the nanoparticles, which does not lead to a proton current; just like in the case of positive external voltages applied to our devices. The found dependence on the type of doping provides strong support for the proposed mechanism. Furthermore, this model does not imply changes in $E$, explains the presence of the photo-effect at all biases and agrees with stable proton currents. It also explains naturally the observed $I \propto P^{1/4}$ dependence. In graphene the electronic temperature depends on illumination power density as[24,27] $T_e \propto P^{1/n}$, with n≥3. Accordingly, we obtain $V_{ph} \propto T_e \propto P^{1/n}$ and, hence, $I \propto P^{1/n}$. In summary, the suggested mechanism is consistent with all our experimental observations.

In terms of applications, the photo-proton effect also deserves attention. The performance of our proton-based devices is even more encouraging if we consider that they present the first, non-optimized design. For example, the addition of photosensitive materials such as quantum dots can be expected to improve their performance[29]. The use of plasmonic nanostructures placed on top of graphene[22,23] may also result in further improvements. Another important application could be in fuel cells and for hydrogen isotope separation[1-4]. Indeed, solar illumination increases the effective proton conductivity of graphene membranes by an order of magnitude at low voltages, which would proportionally reduce internal losses in fuel cells. Membranes for artificial leaves[30] are yet another interesting prospect. The latter application requires some stringent conditions on the membranes including mixed proton-electron conductivity, gas impermeability, mechanical stability and optical transparency[30]. Currently, a mixture of proton and electron conducting polymers is used, but this involves some substantial trade-offs[30] that could be avoided using graphene. Most enticing, however, is perhaps the unexpected richness in properties and phenomena of our system where protons, electrons and photons are packed at an atomically thin interface.

## Methods

**Device fabrication**. Our devices were fabricated by suspending monolayers of mechanically exfoliated graphene over apertures (≈10 μm in diameter) etched into silicon-nitride membranes (see Supplementary Figure 1). On the top side of the devices, graphene was electrically contacted using a microfabricated Au electrode patterned onto the substrate prior to transferring graphene. The membranes were then decorated with a discontinuous layer of Pt or other metals (nominally ∼2 nm) deposited using electron-beam evaporation. The opposite side of the membrane was coated with a Nafion film (5% solution; 1100 EW) and electrically contacted with a proton-injecting electrode ($PdH_x$). The whole assembly was annealed in a humid atmosphere at 130°C to crosslink Nafion.

**Electrical measurements**. Typically, we fixed the voltage bias $V$, measured current $I$ and then light was shined in one-minute ON-OFF pulses. Several such pulses were applied for each voltage and each illumination power density to ensure the reproducibility of our measurements. The light source was a calibrated Newport Oriel Sol3A setup that produced simulated solar illumination of 100 mW $cm^{-2}$. For illumination at lower power densities we used the simulator's aperture diaphragm, which allowed us to control the intensity from 0.7 mW $cm^{-2}$ to 100 mW $cm^{-2}$. We normally performed our electrical measurements in air, unless the response at low biases (~10 mV range) was investigated. In the latter case our devices were placed in a 100% $H_2$ atmosphere at 100% relative humidity. Keithley's SourceMeter was used to both apply $V$ and measure $I$.

**Mass spectrometry.** To measure the hydrogen flow, each device was clamped with O-rings to separate two chambers: one connected to a gas mixture (10% $H_2$ in Ar, 100% humidity) and the other evacuated and connected to a mass spectrometer. The Pt layer faced the vacuum chamber whereas the Nafion layer was in the gas chamber. The graphene membrane served as the cathode (contacted with a microfabricated Au wire; see Supplementary Figure 1) and a dc voltage was applied between graphene and the $PdH_x$ electrode. In these experiments, the gas flow and the electric current were measured simultaneously. As the mass spectrometer, we used an Inficon UL200 Detector.

# Giant photo-effect in proton transport through graphene membranes

Marcelo Lozada-Hidalgo, Sheng Zhang, Sheng Hu, Vasyl Kravets, Francisco J. Rodriguez, Alexey Berdyugin, Alexander Grigorenko, Andre K. Geim

## Device fabrication

The device fabrication procedure is described in the Methods section of the paper and further on ref. [1].

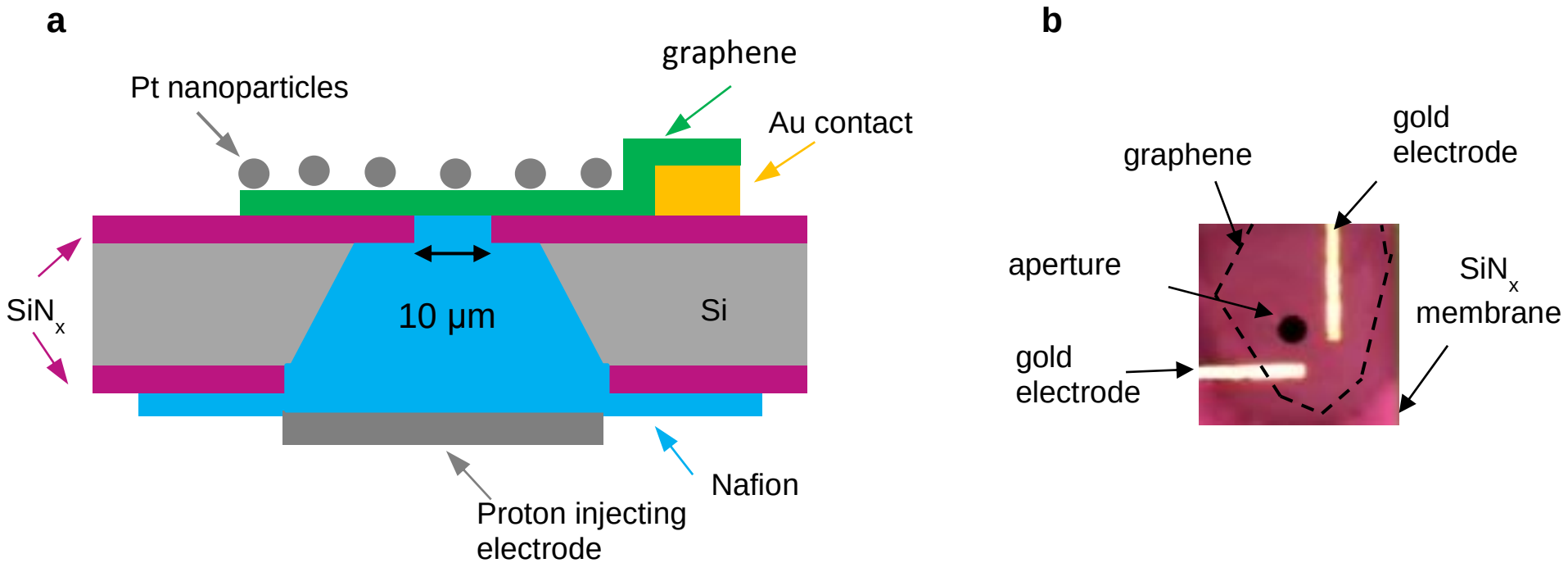

**Supplementary Figure 1|Device geometry. a,** Schematic of our devices. **b**, Optical image of one of our devices. The device shows two gold contacts to the graphene flake that were normally shorted. The dashed lines show the area covered with graphene. The circle in the middle (10 μm diameter) is the aperture in the silicon-nitride membrane.

## Electrical measurements

Note that in our devices the current is not limited by Nafion. Indeed micro-devices normally enable very high current densities. The limiting resistance ($R$) of a cell with a microelectrode of radius $r$ and electrolyte conductivity $\kappa$ is[2,3]: $R = (4\pi\kappa r)^{-1}$. In our case[4], $\kappa \sim 50$ mS cm$^{-1}$, $r \approx 5$ µm and hence we have $R \sim 3$ kΩ. In contrast, the largest current we reported was $I \sim 10$ µA with a bias of ~3V, which translates into a resistance of $R \sim 300$ kΩ. This latter value is ~100 times larger than the limiting cell resistance.

## Additional electrical measurements

To gain further understanding of our devices we performed some additional experiments. We first studied the dark current in devices decorated with metal nanoparticles of Pt, Au, Ni or no metal at all. Changing the metal to Ni or Au reduced the current density for a given voltage bias (Supplementary Fig. 2a), as expected from the known electrochemical activities of these metals towards hydrogen evolution[5]. In contrast, if no nanoparticles were added, no dark or bright current could be detected. This is because protons only transfer across graphene if the opposite side of the membrane is in contact with a material that allows the proton to bond in a stable compound (e.g. Nafion or metal nanoparticles).

We then tested the temperature dependence of the transport in Pt decorated graphene, the best performing devices. In systems that involve electrochemical reactions, it is customary to define a quantity known as the exchange current density[3]. Such quantity is given by: $I_0 = k_B T/(eRA)$, where $A$ is the area of the device, $R$ is the resistance of the devices in the micropolarization region ($\pm$ 5mV), $k_B$ is the Boltzmann constant $T$ is the temperature and $e$ is the elementary charge. In our devices, such current displayed an Arrhenius type behavior: $I_0 \propto \exp(-E_a/k_B T)$, where $E_a$=0.4$\pm$0.06 eV is the activation energy, $k_B$ is the Boltzmann constant $T$ is the temperature. Within the resolution of our measurement, we did not observe any effect of illumination on the activation energy.

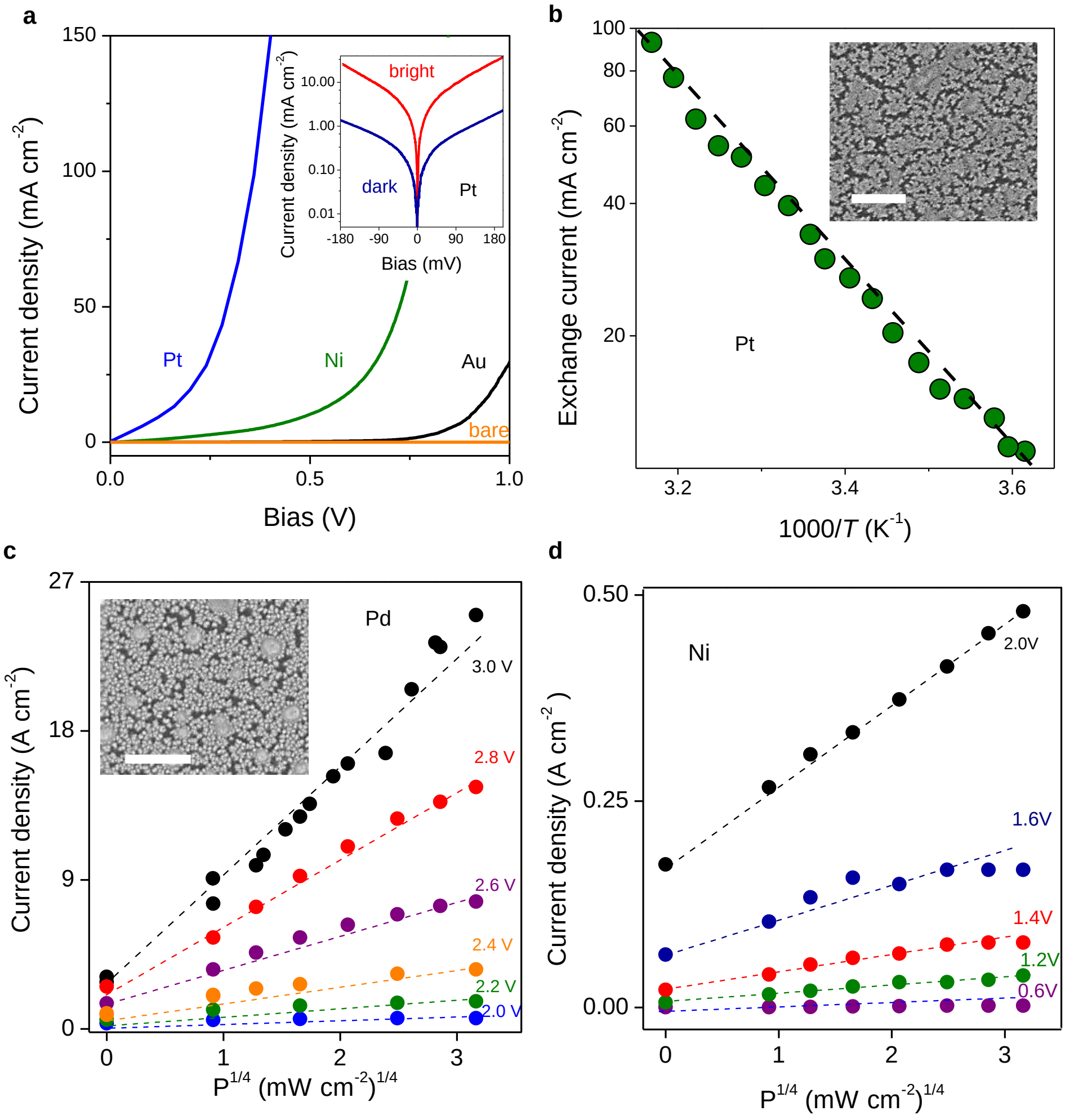

**Supplementary Figure 2|Photoresponse for devices decorated with other metals. a,** *I-V* characteristics in dark for devices decorated with Pt, Ni, Au nanoparticles and for a devices with no nanoparticles at all (bare). Inset, photo-proton effect for Pt in log scale, showing the factor of ~10 scaling between dark and bright response. **b**, Temperature dependence measurements of the exchange current density of Pt decorated devices. Inset, SEM micrograph of graphene decorated with 2 nm of ebeam evaporated Pt. Bright, nanoparticles; dark, graphene. Scale bar 100 nm. **c,d,** Dependence of current on illumination power density for different voltage biases. A power law, $I \propto P^{1/4}$, was observed for devices decorated with Pd (a) and Ni (b). Inset in (c) shows SEM micrograph of graphene decorated with 2 nm of ebeam evaporated Pd. Bright, nanoparticles; dark, graphene. Scale bar 100 nm.

We also observed the photo-proton effect in devices decorated with other metal nanoparticles. Supplementary Figure 2 shows the power dependence of the photoresponse in devices fabricated in the same way as the ones in the main text but with ebeam evaporated Pd or Ni instead of Pt. We observed the same $I \propto P^{1/4}$ response described in the main text for these latter metals; albeit, the photoresponse was a factor of ~2 smaller than with Pt. In contrast, using Au or no metal at all yielded no measurable photoresponse even under maximum illumination power density. These results, then, evidence the importance of the particular metal used to decorate graphene. Furthermore, they rule out a purely plasmonic effect or heating of the device by the illumination as the source of the photo-proton effect.

At this point it is relevant to comment on the structure of the nanoparticles dispersed on graphene. Evaporating metals via ebeam in our membranes results in the formation of nanoparticles a few nanometers wide that aggregate into clusters on the surface of graphene, as can be appreciated in Supplementary Fig. 2b and c (insets). Nanoparticles of Pt, Pd and Au can be expected to contain only an atomically thin native oxide layer. The case of Ni is somewhat different since it oxidizes in air[6]. However, the catalytically active region of these nanoparticles would be the one in direct contact with graphene[5] and, since graphene is impermeable to all gases[6,7], this region would remain protected. It is likely that further protecting theses nanoparticles from oxidation could lead to a higher catalytic activity than the one we observe in our devices[5]. In any case, we note that the interaction of the metal dispersed on the graphene membrane strongly depends on the particular metal[5,8-10]. Experimental and theoretical investigations of adsorbed metals on graphene have shown that Pt, Pd and Ni covalently bond to graphene, whereas Au adsorbs on graphene via van der Waals forces[5,8-10]. Indeed, STM measurements on graphene decorated with Pt nanoparticles showed that forces of several nN cannot detach the nanoparticles from graphene[9]. In contrast, experiments of Au nanoparticles on graphene membranes demonstrated that such particles can be stimulated to move over the surface of graphene just by heating them with a laser[11]. These adhesion properties are important factors in determining how these metals dope graphene and have dramatic consequences in our system, as discussed in the main text.

To gain further insight into the photo-effect, we changed the geometry of our devices such that the graphene membrane was sandwiched in between two layers of Nafion and two proton injecting electrodes. In this case, the photo-proton effect was not observed for any illumination or voltage bias conditions. This result is consistent with the explanation provided in the main text. In this geometry, protons in the Nafion layer that is in contact with the Pt nanoparticles immediately neutralize the electrons sitting on the Pt nanoparticles. This leads to the suppression of the in-plane electric field and, hence, of the photo-effect.

**Spectral dependence of photoresponse**

We studied the spectral dependence of the photoresponse of our devices in the wavelength region 400 nm - 1480 nm. To that end, a broadband laser driven white light source (Energetiq EQ-99X) was used and different wavelength ranges were filtered. An important characteristic of this light source is that the power density is approximately flat over the entire wavelength range measured[12]. This allows for the comparison of the photoresponse of the device over different wavelengths. In the 400 nm - 750 nm range, we used a liquid crystal tunable bandpass filter to isolate individual wavelengths within a ~50 nm range, see Supplementary Figure 3c. Next, we used a short and a long pass filter to isolate the entire 780

nm - 1480 nm range from our white light source. Within this range, to probe into a particular wavelength, we used an additional 1000 nm laser to illuminate the samples.

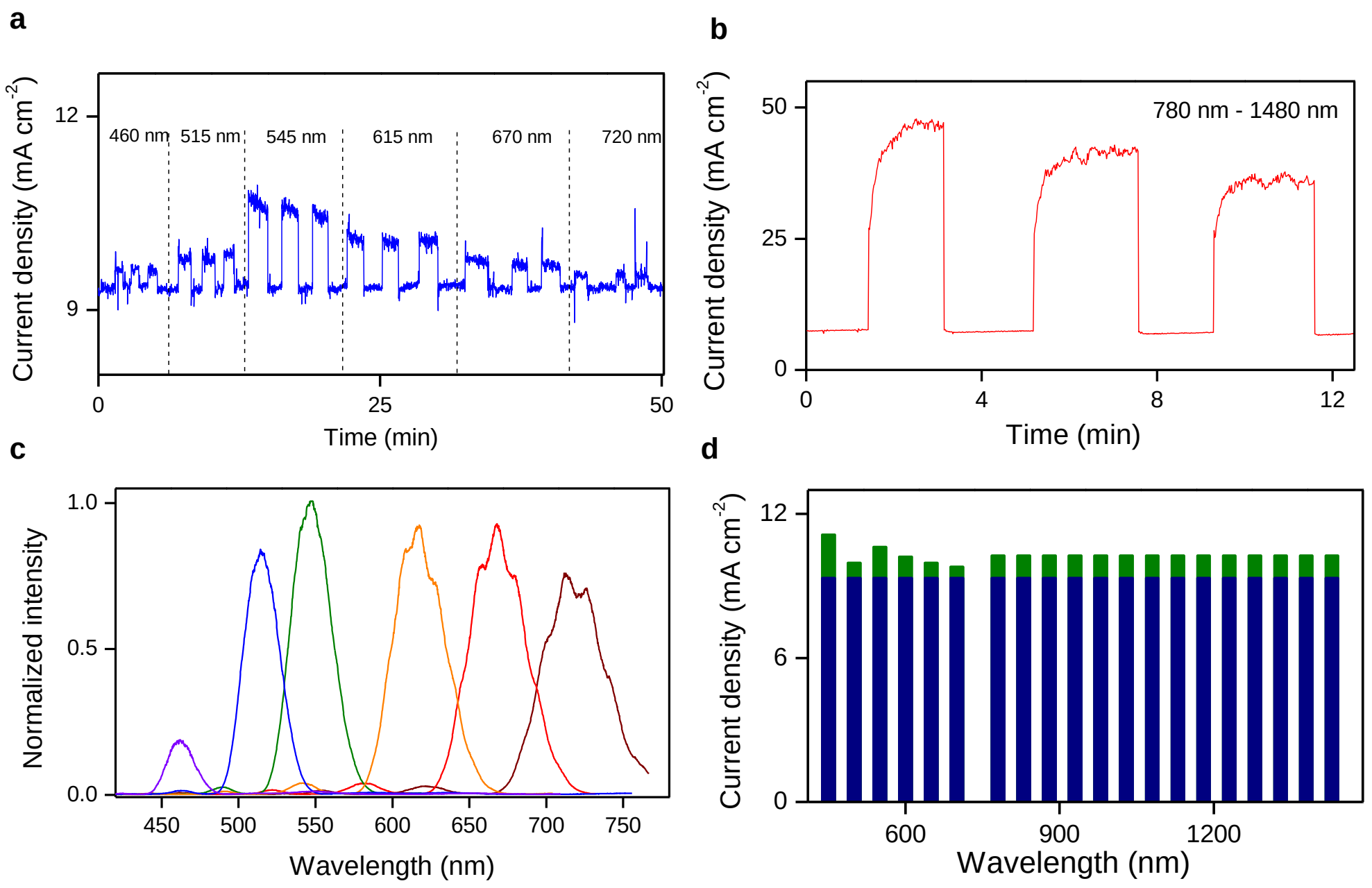

**Supplementary Figure 3|Spectral dependence of photoresponse. a**, Photoresponse for individual wavelengths ranging from 400-800 nm. Applied bias, 1.1 V. Three ON-OFF cycles per wavelength. **b**, Three bright current pulses obtained by shining the whole 780 nm - 1480 nm range. Applied bias, 1.1 V. **c**, Optical fiber measurements of the light filtered by the liquid crystal filter over the wavelengths used in panel a. **d**, Spectral response summary of the data in part a and b. The wavelength range from 780 to 1480 was divided in 50 nm intervals to compare the response observed in the range 450-780. Dark current is shown in blue, bright current in green laser.

The photoresponse of our devices under these conditions is shown in Supplementary Fig. 3 a,b and d. In order to compare the data from these measurements, first, we normalized the data in Supplementary Figure 3a using the relative intensity of the light filtered by the liquid crystal filter (see Supplementary Figure 3c). Next, we divided the observed photoresponse in the 780 nm – 1480 nm between 15 intervals, 50 nm wide. In here, we note that because the devices also showed a response when illuminated with a 1000 nm laser (Supplementary Fig. 3b), the photoresponse in this range is most likely due to contributions from each of the wavelengths involved. The result of this analysis is shown in Supplementary Figure 3e and shows that there is no significant spectral peak in our device's photoresponse over the measured wavelength range.

### Photodetector figures of merit

From our electrical measurements with solar simulated illumination it is possible to extract the photoresponsivity, $R$, of our devices. We extracted this parameter from the shift in current as a function of illumination power density (slope of dotted line in Figure 1b). The extracted value was $R$~$10^4$ A $W^{-1}$. From these data it is also possible to estimate the maximum gain $g$ in our devices through the relation: $g = RP/(e\phi)$ where $e$ is the elementary charge, $\phi$ is the photon flux from our lamp and $P$ is the illumination power density. Hence, from the known photon flux[13] from solar illumination at $P$=100 mW cm$^2$, $\phi \sim 10^{17}$ s$^{-1}$ cm$^{-2}$, we deduce $g \sim 10^4$.

The time response of the photocurrent was measured using a broadband laser driven white light source. The light was mechanically chopped at frequencies up to 1 kHz and focused onto the device. The photocurrent values were derived from the drop in voltage through a 1 kΩ resistor connected in series with the device and the voltage source. Measurements were made with an oscilloscope and carried out with the devices in air. We note that the time response of our devices was limited by the electrical circuit's RC constant. This was shown by using different resistors and observing the corresponding change on the time constant. Thus, we can only place an upper bound to the response time as shown in the main text (≤50 µs). Another consequence of these measurements is that the devices can easily undergo around $10^6$ ON-OFF cycles. Indeed, we measured our devices at 1 kHz for over ten minutes without any sign at all of deterioration.

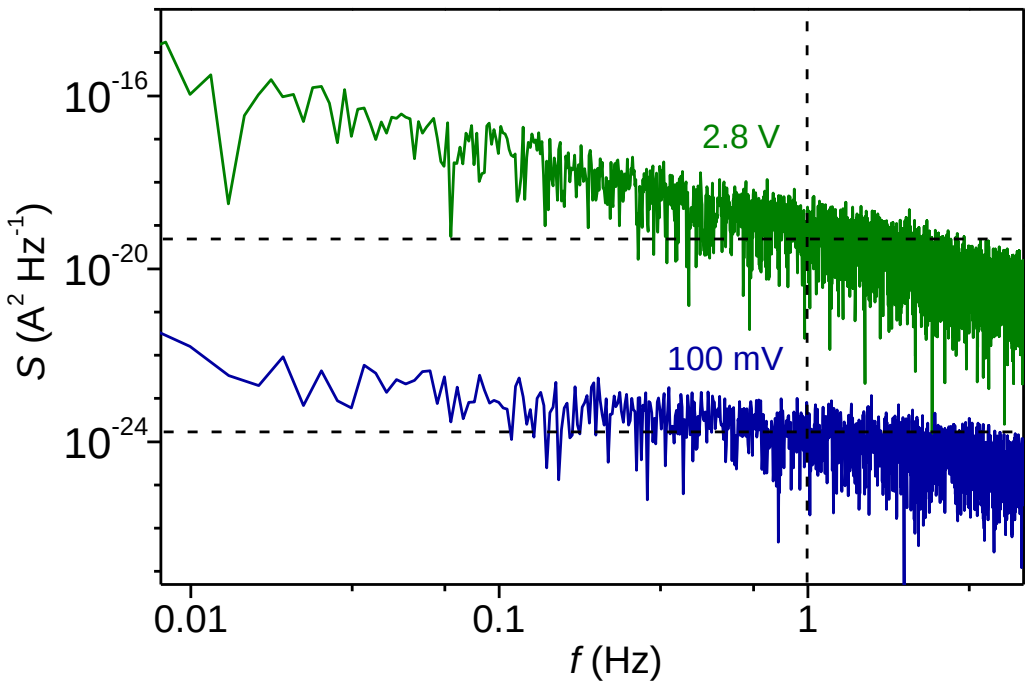

**Supplementary Figure 4|Noise characterization.** Current noise spectral density in the dark at fixed voltage biases.

In order to estimate the noise level of our devices at operational voltages, we measured the current across the device in the dark as a function of time under a fixed voltage bias ranging from 0 V to 2.8 V. To this end, the current was sampled every 100 ms during 5 minutes. Fast Fourier Transform of the measured signal was then applied to find the spectral density of current noise, $S$ (Supplementary Figure 4). From this value, we can extract the noise equivalent power. This parameter measures the radiant power incident on the detector that produces a signal equal to the root mean square detector noise. This quantity is deduced from the spectral noise density as: $NEP=[S(f=1\text{ Hz})]^{1/2} R^{-1}$ , where $S(f=1\text{ Hz})$ is the spectral noise density at 1Hz and $R \sim 10^4$ A W$^{-1}$ is the responsivity of the devices. From our measurements, we deduced a $NEP \sim 1\text{x}10^{-14}$ W Hz$^{-1/2}$ at 2.8 V. The same measurements for small biases, allows us to obtain the noise level floor in our devices and estimate the maximum sensitivity attainable with them. This analysis yielded a $NEP \sim 1\text{x}10^{-16}$ W Hz$^{-1/2}$.